\documentclass[12pt]{article}

\newcommand{\be}{\begin{equation}}
\newcommand{\ee}{\end{equation}}  
\newcommand{\bea}{\begin{eqnarray}}
\newcommand{\eea}{\end{eqnarray}}
\newcommand{\hX}{\hat{X}}
\newcommand{\tr}{\mbox{Tr}}
\newcommand{\str}{\mbox{STr}}
\newcommand{\tg}{\tilde{\gamma}}
\newcommand{\rsl}{r'\!\!\!\!/}

\def\abstract#1{\begin{center}{\large ABSTRACT}\end{center} \par #1}
\def\title#1{\begin{center}{\Large\bf {#1}}\end{center}}
\def\author#1{\begin{center}{\large #1}\end{center}}
\def\address#1{\begin{center}{\it #1}\end{center}}

\newcommand{\EQ}{\begin{equation}}
\newcommand{\EN}{\end{equation}}
\newcommand{\EQA}{\begin{eqnarray}}
\newcommand{\EQN}{\end{eqnarray}}
\newcommand{\EQAN}{\begin{eqnarray*}}
\newcommand{\EQNN}{\end{eqnarray*}}

\newcommand{\Tr}{{\rm Tr}}
\def\identity{{\rlap{1} \hskip 1.6pt \hbox{1}}}

\makeatletter
\@addtoreset{equation}{section}
\makeatother
 
\begin{document}
\begin{titlepage}
\hspace*{\fill}
\vbox{
\hbox{hep-th/0201248}
\hbox{UT-02-03}
\hbox{TIT/HEP-474}
\hbox{January, 2002}}
\vspace*{\fill}
\begin{center}
  \Large\bf 
Non-abelian action of D0-branes from
 Matrix theory \\
in the longitudinal 5-brane background
\end{center}
\vskip 1cm
\author{
Masako Asano \footnote{E-mail address:\ \ 
{\tt asano@hep-th.phys.s.u-tokyo.ac.jp}}}
\address{
Department of Physics,  University of Tokyo, \\
 Hongo, Bunkyo-ku, Tokyo 113-0033}
 
\begin{center} and \end{center}

\author{
Yasuhiro Sekino \footnote{ 
E-mail address:\ \ {\tt sekino@th.phys.titech.ac.jp}}}
\address{
Department of Physics,  Tokyo Institute of Technology, \\
Ookayama, Meguro-ku, Tokyo 152-8550}

\vspace{\fill}
\abstract{We study one-loop effective action of Berkooz-Douglas
Matrix theory and obtain non-abelian action of D0-branes
in the background field produced by longitudinal 5-branes.
Since these 5-branes do not have D0-brane charge
and are not present in BFSS Matrix theory, our analysis
provides an independent test for the coupling 
of D-branes to general weak backgrounds proposed  
by Taylor and Van Raamsdonk from the analysis of the BFSS
model.
The proposed couplings appear  
in the Berkooz-Douglas effective action precisely
as expected,
which suggests the consistency of the two matrix models.
We also point out the existence of the terms which are not given 
by the symmetrized trace prescription in the Matrix theory
effective action.}

\vspace*{\fill}


\end{titlepage}

\section{Introduction}
D-branes \cite{P} have played crucial roles in 
understanding  the string dualities, 
as well as in attempting to give a formulation of M-theory.
In weakly-coupled string theory, D-branes appear as solitonic
objects which allow a remarkably simple description:
D$p$-branes are defined as  $(p+1)$-dimensional 
hypersurfaces which support endpoints of open strings.
The fact that D-branes couple to gravity via
open-string closed-string interaction suggests that 
they must be considered as dynamical objects. 
The open-string massless scalar modes  which live
on the $(p+1)$-dimensional world-volume corresponds
to the collective coordinates of D$p$-branes.
A notable feature of D-branes is that when $N$ D-branes 
are coincident, the transverse motion is described 
by $N\times N$  matrices rather than just $N$ coordinates, 
due to the presence of extra massless scalars coming from 
the strings connecting different D-branes \cite{W}. 
Understanding the dynamics of D-branes and especially,
revealing the consequence of the non-commutativity 
of matrix-valued coordinates are undoubtedly
important for the further clarification
of the non-perturbative nature of string theory.

Effective action for a single D-brane is well-understood.
A D$p$-brane is described by the ($p+1$)-dimensional 
Born-Infeld (BI) action plus Chern-Simons (CS) terms, 
in the low-acceleration limit (where the second
derivatives of the fields are neglected).
The $(9-p)$ scalars describe the position of the brane and 
there are gauge fields corresponding to the U(1) symmetry.
BI action is obtained from the condition of the conformal 
invariance on the string world-sheet and includes 
all the $\alpha'$ corrections associated to the open-string
massless modes \cite{Le}.
CS terms give the coupling of a D$p$-brane to Ramond-Ramond $(p+1)$-form 
potential and also to lower ($p-1$, $p-3$, \ldots)-form 
potentials in the presence
of non-trivial configuration of U(1) gauge fields \cite{Li, D1}.

For the action of multiple D-branes, 
we have only limited understanding at present.
The leading terms of the low-energy effective action of
$N$ D$p$-branes in flat space is given by the 
$D=(p+1)$ U($N$) supersymmetric Yang-Mills theory which
is obtained by the dimensional reduction from the 10D SYM \cite{W}.
Contrary to the case of a single D-brane, 
non-commutativity of field strengths makes it difficult to obtain
effective action to all orders in $\alpha'$, even in flat space. 
It was argued by Tseytlin that the part of the action 
independent of the commutator of the field strengths
is given by a non-abelian generalization of BI action in 
which the trace 
for the gauge group is taken after symmetrizing the product \cite{T1}.
However, full form of the action is not understood.
The contribution at each order in $\alpha'$ should be determined 
from the analysis of scattering amplitude 
of the open-string massless states, as in refs.\cite{GW, T2}.
Indeed, there are suggestions that there must be corrections
to the Tseytlin's action at the sixth order of field 
strengths \cite{HT}.

How to couple multiple D-branes to curved background is 
further unclear. 
As mentioned above, the collective coordinates of D-branes
are promoted to matrices, thus the background geometry should
be regarded as a function of matrices. 
Principle for writing an action in such circumstances
is obscure, despite some attempts \cite{D2}.
In ref.\cite{dB}, a generalization of the notion of the general
coordinate transformation is discussed, but the constraint from 
that symmetry does not seem to be enough for determining the 
action unambiguously. 
An approach taken in refs.\cite{My,TR1} is to
treat the background fields 
as an expansion around a point in spacetime, in which 
the coordinates in the expansion are replaced by matrices
(`non-abelian Taylor expansion'). 
Scattering amplitudes of a closed-string state
and open-string states studied in refs.\cite{GM1,GM2} 
give some evidence for the consistency of 
the above approach, but the form of the action at
 higher orders of the expansion, and especially, 
how to order the matrices are not clear.
Important observation about the action of multiple D-branes
is that non-abelian version of CS terms should allow the coupling
of D$p$-branes to RR forms which are of higher degree than 
$(p+1)$-forms by non-commutative configurations of scalar 
fields \cite{My,TR1}.

Matrix theory \cite{BFSS}, which is the proposal for the
exact definition of M-theory in the light-cone frame,
provides an alternative way to study interactions of
D-branes. 
In a series of papers 
by Kabat and Taylor \cite{KT} and 
by Taylor and Van Raamsdonk \cite{TR1,TR2,TR3},
detailed study of Matrix theory effective action
was performed. 
In refs.\cite{KT,TR1}, 
by reinterpreting the one-loop effective potential of Matrix theory
 as the result of tree-level supergravity interactions,
 Matrix theory operators which couple to the supergravity fields
were identified. Taylor and Van Raamsdonk 
proposed the Matrix theory action in general weak 
backgrounds using those couplings \cite{TR1}. 
Further, D0-brane action in weak 10D background
was obtained \cite{TR2}, 
following the scaling argument due to Seiberg \cite{Sei} 
and Sen \cite{Sen} which relates Matrix theory to 10D
type IIA string theory. 
Applying T-duality, similar couplings for general D$p$-brane were
given in ref.\cite{TR3}. 
Consistency of the couplings 
which were obtained in this way has been  confirmed in several
contexts: the D9-brane action is indeed 10-dimensional Lorentz 
covariant \cite{TR3}; 
the coupling of D0-branes to the background field 
is consistent with the ones obtained from the matrix-regularization
of the supermembrane in a curved background \cite{DNP}; and
absorption cross sections of dilaton partial waves by D3-branes
which were evaluated by semi-classical gravity
are reproduced by the gauge theory using the above couplings
\cite{KTR}.

In this paper, we study a variant of Matrix theory which
was proposed by Berkooz and Douglas as the definition of M-theory
in the light-cone frame {\it in the presence of 
longitudinal 5-branes} \cite{BD}.
As reviewed in section 2,
Berkooz-Douglas (BD) Matrix theory has extra degrees of 
freedom compared to 
the original Matrix theory. 
We integrate them out at
one-loop order and obtain effective action for the 
D0-brane degrees of freedom in the background fields
produced by the longitudinal 5-branes. 
Since the 5-branes which we are discussing
have no D0-brane charge, they are not present in the ordinary
Matrix theory. Thus, our analysis can be regarded as
an independent test of the Taylor and Van Raamsdonk's proposal for
the D-brane action in weak background fields, and 
also as a check of the consistency between the two formulations of
Matrix theory. 
We confirm that the couplings expected from the above
proposal indeed exist in the effective action of BD
Matrix theory. In addition to the proposed couplings,
we find corrections involving
extra commutators in the Matrix theory effective action.
We also discuss the consistency of the effective action
of BD Matrix theory and a proposal of Myers \cite{My}
for the D-brane action in curved space.

This paper is organized as follows. In section 2, we 
review BD Matrix theory and set notations. 
In section 3, one-loop integration of 
 the massive fields is performed. In section 4, 
we compare the effective action obtained
in section 3 with Taylor and Van Raamsdonk's proposal for 
the Matrix theory action in weak background fields.
In section 5, we comment on the interpretation of
our result from the perspective of 10D string theory.
In section 6, we conclude and discuss directions
for the future works.

\section{Matrix theory in the longitudinal 5-brane background}
According to the Matrix-theory conjecture of Banks, Fischler,
Shenker and Susskind (BFSS) \cite{BFSS}, 
M-theory in the infinite momentum frame (IMF)
is defined by the large $N$ limit of 
the supersymmetric quantum mechanics with U($N$) gauge
symmetry, which is the effective action of $N$ D0-branes
in the low-energy limit. 
A D0-brane has a unit positive momentum in the longitudinal (11-th)
direction and is a natural candidate for the basic constituent
in the IMF.
In ref.\cite{BD}, Berkooz and Douglas proposed a formulation
of M-theory in the presence of longitudinal 5-branes.
Longitudinal 5-branes, which fill the 11-th direction and have
 zero longitudinal momentum in their ground state, are considered
as non-trivial background in the IMF. 
Note that the `5-brane in the ground state' 
does not have D0-brane charge and cannot be constructed 
in ordinary Matrix theory.
Based on the philosophy that different vacua give rise
to different Hamiltonians in the IMF in general,
modification of Matrix theory was conjectured.
Compared to the original BFSS Matrix theory, this theory
has extra degrees of freedom, and has only half of 
the supersymmetries.

Precisely, the action of Berkooz-Douglas (BD) Matrix theory 
is the 0-0 and 0-4 string sectors of the SYM describing 
the D0-D4 bound state, which is given by the dimensional
reduction of the D=6, ${\mathcal N}=1$ SYM.  
In the case of $N$ D0-branes and $N_4$ D4-branes,
the 0-0 sector fields, which are the degrees of freedom
of the original Matrix theory, are in adjoint rep. of U($N$). 
The 0-4 sector fields are the hypermultiplets of the 6D theory
which consist of bosons with 4 real components  
and fermions with 8 real components, both of which
transform as the bi-fundamental rep. of 
$U(N)\times U(N_4)$.
We consider the  BD Matrix theory as M-theory 
in the presence of $N_4$ longitudinal 5-branes, which is
compactified in the light-like direction $x^-$
with total longitudinal momentum $P_-=N/R$ 
(where $R=g_s\ell_s$), following the usual DLCQ interpretation 
of finite $N$ Matrix theory \cite{Su}.

The action is given as follows.
\begin{equation}
S=S_0+S_5
\label{BDaction}
\end{equation}
\begin{eqnarray}
S_0&=& {1\over g_s\ell_s}\int dt \Tr \Big( {1\over 2}D_0 X_i D_0 X_i
+{1\over 4 \lambda^2} [X_i,X_j]^2 
+{i\over 2}\overline{\Theta}\Gamma^0 D_0\Theta 
+{1\over 2\lambda}\overline{\Theta}\Gamma^i [\Theta, X_i] \Big)
\label{S0}\\
S_5&=&\int dt \Big\{ (D_0 v_I)^\dagger D_0 v_I 
-{1\over \lambda^2}v_I^\dagger (X_a -Y_a)^2 v_I -i\chi^\dagger D_0 \chi 
-{1\over \lambda}\chi^\dagger \gamma^0\gamma^a(X_a -Y_a)\chi\nonumber\\
&& -{1\over 2\lambda^2}\Big( v_1^\dagger ([\phi_1,\bar{\phi}{}_1] 
+[\phi_2,\bar{\phi}{}_2])
v_1 - v_2^\dagger ([\phi_1,\bar{\phi}{}_1] +[\phi_2,\bar{\phi}{}_2])
v_2 \nonumber\\
&&-2v_2^\dagger ([\bar{\phi}{}_1,\bar{\phi}{}_2]) v_1
+2v_1^\dagger ([\phi_1, \phi_2]) v_2 \Big)
+(v^4 \mbox{-terms}) +(v\Theta \chi \mbox{-terms})\Big\}
\label{S5}
\end{eqnarray}
where $S_0$ is the part
containing only the 0-0 sector, 
which is the same as the BFSS action, and
$S_5$ is the additional part containing 0-4 sector.
 
Let us explain the notations and conventions. 
We use the indices $i,j =1,\ldots,9$ for the spatial directions 
in the 10D; $m,n =1,\ldots,4$ for the spatial directions tangent to the
5-branes except for $x^{10}$ (i.e. tangent to the D4-brane);
$a,b=5,\ldots,9$ for the directions transverse to the 5-branes.
Length scale is given by $\lambda = 2\pi \ell_s^2$.  
The D0-brane fields $X_i$ and $\Theta$ are 
$N\times N$ Hermitian matrices where $\Theta$ satisfy
the 10D Majorana-Weyl condition.
Covariant derivatives for these fields are defined as 
$D_0 X_i =\partial_0 X_i +i[A_0,X_i]$.
We also use complex combinations of $X_m$ which are defined as
$(\phi_1,\phi_2)=(X_1+iX_2,X_3+iX_4)$ with 
$\bar{\phi}{}_1=\phi_1^\dagger$ and $\bar{\phi}{}_2=\phi_2^\dagger$. 
The 0-4 sector fields are complex bosons $v_I$ ($I=1,2$)  and 
complex fermions $\chi$ which satisfy the 6D Weyl condition
($\bar{\gamma}\chi=\chi$ where 
$\bar{\gamma}\equiv \gamma^0\gamma^5\ldots\gamma^9$) 
Covariant derivatives for the 0-4 fields are
defined as $D_0 v_I =\partial_0 v_I +iA_0 v_I$.
We will put the indices for the bi-fundamental rep. of $U(N)\times U(N_4)$
as $v_I^{A\tilde{A}}$ and $\chi^{A\tilde{A}}$
($A=1,\ldots, N$ and $\tilde{A}=1,\ldots,N_4$) when necessary.
$Y_a$ are $N_4\times N_4$ matrices, {\it i.e.} singlets under
U($N$), which specifies the positions of the 5-branes. 
When 5-branes are coincident, which is the case treated
in this paper, $Y_a$ is proportional to identity matrix 
$\identity_{N_4\times N_4}$. In this case, $Y_a$ can be absorbed into
the definition of $X_a$, so we set $Y_a=0$ hereafter.

Note that we have not adopted a convention using 
the SU(2) Majorana spinors, which may be familiar 
in the literatures (such as refs.\cite{BD,Ra}). 
It is because we prefer unconstrained complex 
spinors to perform the loop calculations.
The fact that SO(4) symmetry in the $X_m$ 
direction is not manifest
in the above expressions is a consequence of that choice, 
but the result of the loop calculation
can of course be written in SO(4) covariant way. 
Also note that we have explicitly written only the part of the
action which is needed for the one-loop integration 
of $v$ and $\chi$.
There are also the $v^4$-terms and the $v\Theta\chi$-terms.
The $v^4$-terms are proportional to $g_s$ in our normalization,
and give rise to  higher-loop corrections.
Half of the components of $\Theta$ (which have definite 6D 
chirality $\bar{\gamma}\Theta=-\Theta$) appear in the 
$v\Theta\chi$-terms.
The action (\ref{BDaction}) is invariant under 
the SUSY transformation
with a 6D Weyl spinor parameter ($\bar{\gamma}\eta=-\eta$), 
and the number of real supercharges is eight.


\section{One-loop effective action}\label{sect3}
\subsection{Method for the perturbative calculation}
In this section, we calculate effective action of the 
D0-brane degrees of freedom $X_i$ in BD Matrix theory
by integrating out $v$ and $\chi$ in eq.(\ref{S5}) 
at one-loop order.
We use Euclidean version of the action 
by transforming $t\rightarrow -i\tau$,
$A_0 \rightarrow iX_0/\lambda$ and $S\rightarrow -i S$. 
We evaluate the one-loop determinant
\begin{equation}
S_{\mbox{\scriptsize eff}} = S_0 - \delta^{(1)},
\end{equation}
\begin{equation}
\delta^{(1)}=\int d\tau \left[- \ln \mbox{Det}\,K_{\mbox{\scriptsize bos}} + 
\ln \mbox{Det}\,K_{\mbox{\scriptsize fermi}} \right]
\label{defeff_action}
\end{equation}
where $K_{\mbox{\scriptsize bos}}$ and $K_{\mbox{\scriptsize fermi}}$ are kernels of
quadratic terms of complex bosons $v$ and fermions $\chi$,
respectively.

In this paper, we take the matrix background as
\begin{eqnarray}
\Theta&=&0\nonumber\\
(X_0,X_m, X_a)& =& (\hX_0,\hX_m ,r_a+\hX_a)
\end{eqnarray}
where 
$\hX_i$ are general time dependent matrices 
and $r_a$ are constants proportional to the
identity matrix.
We divide the part of the action which is quadratic 
in $v$ and $\chi$ 
into free and interaction part  
as follows
\begin{eqnarray}
S_{\mbox{\scriptsize free}}&=& \int d\tau 
\left\{
v^{\dagger}
\left(-\partial_{\tau}^2+\frac{r^2}{\lambda^2}
\right)v +
\chi^{\dagger}\left(-\partial_{\tau}+\frac{1}{\lambda}
\tilde{\gamma}^a r_a
\right)\chi
\right\}
\\
S_{\mbox{\scriptsize int}} &=& \int d\tau \left\{
\frac{1}{\lambda^2}v_I^\dagger (2r_a \hX_a +\hX_a^2 
+\hX_0^2-i\lambda\partial_\tau \hX_0-2i\lambda \hX_0\partial_\tau)v_I
\right.
\nonumber\\
&& 
+\frac{1}{2\lambda^2}\left[v_1^\dagger([\phi_1,\bar\phi_1]
+[\phi_2,\bar\phi_2])v_1 
-v_2^\dagger([\phi_1,\bar\phi_1]+[\phi_2,\bar\phi_2])v_2 
\right.
\nonumber\\
&&
\left. \left. -2 v_2^\dagger[\bar\phi_1,\bar\phi_2]v_1
+2 v_1^\dagger[\phi_1,\phi_2]v_2
\right]
+\frac{1}{\lambda}\chi^\dagger (\tilde{\gamma}^a\hX_a -i \hX_0)\chi 
\right\}
\label{Sint1}
\\
&\equiv& \int d\tau \left\{\frac{1}{\lambda^2}v_I^\dagger V_{IJ}(\tau)v_J 
+\frac{1}{\lambda}\chi^\dagger (\tilde{\gamma}^a\hX_a -i \hX_0)\chi\right\} .
\label{Sint2}
\end{eqnarray}
Here we take $\tilde{\gamma}^a=\gamma^0\gamma^a$  as $4\times 4$ matrices
acting on 6D Weyl spinors ($\overline{\gamma}\chi=\chi$) which
have 4 complex components. Also note
\begin{equation}
\tilde{\gamma}^{a_1a_2a_3a_4a_5}
=\overline{\gamma}\epsilon_{a_1a_2a_3a_4a_5}
=\epsilon_{a_1a_2a_3a_4a_5}
\end{equation}
with $\epsilon_{56789}=1$. 

We adopt a method of calculation which is conceptually most straightforward:
we evaluate one-loop diagrams with suitable number of
vertex insertions, treating the vertices as an 
expansion in derivatives.\footnote{Another method
is to reproduce the effective action from the
Eikonal phase shift. (See e.g. ref.\cite{DKPS})
However, applying this method by 
taking the background $X_5=r$, $X_6=v\tau$ (where 
$r,v\propto\identity$), we could not obtain all the terms 
which are expected in the effective action.
Especially, `Chern-Simons couplings' given in section 3.4
is missing.}
Our method closely follows that of ref.\cite{TR1}
where the one-loop integration of off-diagonal 
blocks in BFSS Matrix theory is performed.
We will set $\hX_0=0$ in all the expressions in the following. 
Since our calculation preserves gauge invariance, 
we can recover the dependence on $\hX_0$ ($A_0$) by 
simply replacing $\partial_\tau$ with $D_\tau$ in the result.

First, we consider contribution from bosons $v$ 
to the effective action.
Propagators are determined from $S_{\mbox{\scriptsize free}}$ as
\[
\left(
-\partial_\tau^2+\frac{1}{\lambda^2}r^2
\right)
\langle v_{I,A\tilde{A}}(\tau) v_{J,B\tilde{B}}^\dagger (\tau') \rangle
=\delta^{IJ}\delta^{AB}\delta^{\tilde{A}\tilde{B}}\delta(\tau-\tau'),
\]
\begin{equation}
\langle v_{I,A\tilde{A}}(\tau) v_{J,B\tilde{B}}^\dagger
(\tau') \rangle
=\delta^{IJ}\delta^{AB}\delta^{\tilde{A}\tilde{B}}
\int\frac{dk}{2\pi}\frac{e^{ik(\tau-\tau')}}{k^2+r^2/\lambda^2}
\equiv \delta^{IJ}
\delta^{AB}\delta^{\tilde{A}\tilde{B}}
\Delta(\tau-\tau').
\end{equation}
The bosonic part of  $\delta^{(1)}$ is
given as
\begin{eqnarray}
\delta^{\mbox{\scriptsize bos}} &=& - \ln \mbox{Det} (K_{\mbox{\scriptsize bos}}^{\mbox{\scriptsize (free)}}+V)
\nonumber\\
& = & \Gamma^{\mbox{\scriptsize bos}}-\ln \mbox{Det} 
K_{\mbox{\scriptsize bos}}^{\mbox{\scriptsize (free)}}
\end{eqnarray}
where
\begin{eqnarray}
\Gamma^{\mbox{\scriptsize bos}} &=&
\tr\sum_{n=1}^{\infty} \frac{(-1)^n}{n} [V(K_{\mbox{\scriptsize 
bos}}^{\mbox{\scriptsize (free)}})^{-1}]^n 
\nonumber\\
&=&  \sum_{n=1}^{\infty} \frac{(-1)^n}{n} \frac{1}{\lambda^{2n}}
\int d\tau_1 \cdots d\tau_n 
\tr [V_{I_1 I_2}(\tau_1) V_{I_2 I_3}(\tau_2)
\cdots V_{I_n I_1}(\tau_n)]
\nonumber\\
&& \quad \times
\Delta(\tau_1-\tau_2) \Delta(\tau_2-\tau_3)\cdots \Delta(\tau_n-\tau_1)
\nonumber\\
&=&  \sum_{n=1}^{\infty} \frac{(-1)^n}{n}\frac{1}{\lambda^{2n}}
\sum_{D_i=0}^{\infty}
\int d\tau \tr \left[V_{I_1 I_2}(\tau) V_{I_2 I_3}^{(D_2)}(\tau)
\cdots V_{I_n I_1}^{(D_n)}(\tau)\right]
\nonumber\\
&& \times
\int \prod_{i=1}^{n} \frac{dk_i}{2\pi}
\int \prod_{i=2}^{n} \left(d\sigma_i 
\frac{\sigma_i^{D_i}}{D_i !}\right)
\frac{e^{-i k_1 \sigma_2}}{k_1^2+r^2/\lambda^2}
\frac{e^{-i k_2(\sigma_2-\sigma_3)}}{k_2^2+r^2/\lambda^2}
\cdots
\frac{e^{i k_n \sigma_n}}{k_n^2+r^2/\lambda^2}.
\end{eqnarray}
To obtain the last expression, we rewrote the vertices
$V_{I_iI_{i+1}}(\tau_1)$ 
using Taylor expansion around a reference point $\tau_1$. 
The superscript $(D_i)$ means the $D_i$-th derivative 
in $\tau$ and $\sigma_i\equiv\tau_i-\tau_1$.
Performing the integration 
over $\sigma_i$ and $k_i$ ($i=2,\cdots, n$), 
$\Gamma^{\mbox{\scriptsize bos}}$ reads
\begin{eqnarray}
\Gamma^{\mbox{\scriptsize bos}} &=&  \sum_{n=1}^{\infty} \frac{(-1)^n}{n}
\sum_{D_i=0}^{\infty} \left(\prod_{i=2}^{n} \frac{1}{D_i !}\right)
\int \!d\tau \, \tr \left[V_{I_1 I_2}(\tau) V_{I_2 I_3}^{(D_2)}(\tau)
\cdots V_{I_n I_1}^{(D_n)}(\tau)\right]
\nonumber\\
&& \hspace{-1cm}\times
\frac{\lambda^{D-1}}{r^{2n+D-1}}
\int\frac{dk}{2\pi}
\frac{1}{k^2+1}(i\partial_k)^{D_2}\!
\left\{
\frac{1}{k^2+1}\!\left[(i\partial_k)^{D_3}\! \frac{1}{k^2+1}\!\left(
\cdots (i\partial_k)^{D_n}\!\frac{1}{k^2+1} \right)\right]
\right\}
\label{gammab}
\end{eqnarray}
where $D=\sum D_i$.
Note that the terms with 
odd number of derivatives do not contribute to 
$\Gamma^{\mbox{\scriptsize bos}}$,
for they are proportional to $\int dk k^{2n+1}/(k^2+1)^{l}$
which are vanishing.

The fermionic propagator is given by
\begin{eqnarray}
\langle \chi_{\alpha,A\tilde{A}}(\tau) 
\chi_{\beta,B\tilde{B}}^\dagger
(\tau') \rangle
&=& \delta^{AB}\delta^{\tilde{A}\tilde{B}}
\int\frac{dk}{2\pi}\frac{e^{ik(\tau-\tau')}}{k^2+r'^2}
\left(r'\!\!\!\!/+ik
\right)_{\alpha\beta}\nonumber\\
&=& \delta^{AB}\delta^{\tilde{A}\tilde{B}}
\left(\partial_\tau+r'\!\!\!\!/
\right)_{\alpha\beta}
\Delta(\tau-\tau')
\end{eqnarray}
where $r'=r/\lambda$ and $ r'\!\!\!\!/=\tg^a r_a/\lambda$.
Contribution from the fermionic loop to the 
effective action is obtained in the same way as
in the bosonic case.
\begin{eqnarray}
\delta^{\mbox{\scriptsize fermi}} &=& \ln \mbox{Det} 
(K_{\mbox{\scriptsize fermi}}^{\mbox{\scriptsize (free)}}+V)
\nonumber\\
& = & \Gamma^{\mbox{\scriptsize fermi}}+\ln \mbox{Det} 
K_{\mbox{\scriptsize fermi}}^{\mbox{\scriptsize (free)}}
\end{eqnarray}
\begin{eqnarray}
\Gamma^{\mbox{\scriptsize fermi}} &=& - \sum_{n=1}^{\infty} \frac{(-1)^n}{n}\frac{1}{\lambda^n}
\sum_{D_i=0}^{\infty} \left(\prod_{i=2}^{n} \frac{1}{D_i !}\right)
\int \!d\tau \, \tr \left[\hX_{a_1}(\tau) \hX_{a_2}^{(D_2)}(\tau)
\cdots \hX_{a_n}^{(D_n)}(\tau)\right]
\nonumber\\ 
&& \hspace*{-2.5cm}\!\!\!\!\times
\int\frac{dk}{2\pi}
\tr\left\{
\tg^{a_1}
\frac{\rsl+ik}{k^2+r'^2}\tg^{a_2} (i\partial_k)^{D_2}\!
\left[
\frac{\rsl+ik}{k^2+r'^2}\tg^{a_3}
(i\partial_k)^{D_3}\!\left(\cdots 
\tg^{a_n}(i\partial_k)^{D_n}\!\frac{\rsl+ik}{k^2+r'^2} )\right)
\right]
\right\}.
\label{gammaf}
\end{eqnarray}
where  the trace in the first line of eq.(\ref{gammaf})
is for gauge indices
and the one in the second line is for spinor indices.

As a consequence of the supersymmetry, 
one-loop determinant of the free 
propagator of bosons and fermions cancel each other
$(-\ln \mbox{Det} K_{\mbox{\scriptsize bos}}^{\mbox{\scriptsize (free)}}+\ln \mbox{Det} K_{\mbox{\scriptsize fermi}}^{\mbox{\scriptsize (free)}}=0)$,
thus the 
effective action is given by $\delta^{(1)}=\Gamma
=\Gamma^{\mbox{\scriptsize bos}}
+\Gamma^{\mbox{\scriptsize fermi}}$.
We obtain effective action in the expansion with respect
to the number of vertices and of derivatives.

The region where our expansion is good is
when D0-branes are slowly moving
and nearly coincident, as explained below.
Firstly, the expansion in derivatives
is justified when $\lambda\partial_t{\hX_i}/r^2$ is small.
It is because each derivative is associated with a
factor $\lambda/r$ and each $\hX_i$ is associated with
$1/r$, as we see from eqs.(\ref{gammab}) and (\ref{gammaf}).
The expansion in $\hX_i$ is good when $\hX_i/r$ is small.
The one-loop approximation is justified when $g_s 
\lambda^{3/2}/r^3$ is small: As we go to one higher loop,
$v^4$ vertex is inserted once which contributes a factor $g_s$,
and two extra propagators and one extra momentum integral
are needed which contribute a factor 
$\lambda^{3/2}/r^3$.\footnote{These conditions are the same as 
the one for the perturbation in BFSS Matrix theory \cite{BBPT}.}

In the following, we present the result of the calculation
for the terms containing up to two derivatives. 
Explicitly, up to fourth order 
in $\hX_i$ for $D=2$ and to sixth order in $\hX_i$ for $D=0,1$.
We denote by $\Gamma((\hX_a)^{N_a} (\hX_m)^{N_m},D)$ 
the term which contains $N_a$ $\hX_a$'s and ${N_m}$ $\hX_m$'s 
and $D$ derivatives.

\subsection{Potential terms}
First, we consider terms with no derivatives.
For the terms with only $(\hX_a)$'s, we have
\begin{eqnarray}
\Gamma((\hX_a)^{N_a},D=0 ) &=& 
\frac{N_4}{8\lambda}\frac{1}{r^3}
\int d\tau \str [\hX_{a_1},\hX_{a_2}]^2
\label{pota0}
\\
&&-\frac{3N_4}{8\lambda}\frac{r_a}{r^5}
\int d\tau \str (\hX_a\,[\hX_{a_1},\hX_{a_2}]^2)
\label{pota1}
\\
& &-\frac{3N_4}{16\lambda}\frac{1}{r^5}
\int d\tau \tr ((\hX_{a})^2[\hX_{a_1},\hX_{a_2}]^2) 
\label{pota2A}
\\ && +
\frac{15N_4}{16 \lambda }\frac{r_{a_1}r_{a_2}}{r^7}
\int d\tau \str ( \hX_{a_1}\hX_{a_2}[\hX_{b_1},\hX_{b_2}]^2 )
\label{pota2B}
\\
& &+
\frac{N_4}{8 \lambda}\frac{1}{r^5}
\int d\tau \tr ( [\hX_{a_1},\hX_{a_2}][\hX_{a_2},\hX_{a_3}] [\hX_{a_3},\hX_{a_1}]) 
\label{potac}
\\&&
\quad +O((\hX_a)^7 )\nonumber
\end{eqnarray}
where $\str (\cdots)$ stands for symmetrized trace,
which means that trace operation is taken after symmetrizing 
the ordering of all
$[\hX_{a_1},\hX_{a_2} ]$, $\dot{\hX}_a$
 and $\hX_a$ 
in the parenthesis.
Note that for the first two lines of the above equation,
there is no difference between $\tr$ and $\str$.
Vanishing of the terms containing fewer number of 
$X_a$'s can easily be proved.

Now we consider the terms containing $\hX_m$. 
As we see from (\ref{Sint1}), $\hX_m$ appear only in 
the vertex for bosons which is 
of the form  $\sim v[\hX_{m_1},\hX_{m_2}]v$. 
Thus, 
\begin{equation}
\Gamma((\hX_m)^{2N_m+1} (\hX_a)^{N_a}, D)=0 
\label{Xmrelation1}
\end{equation}
holds generally. Also, 
\begin{equation}
\Gamma((\hX_{m})^2(\hX_{a})^{N_a},D)=0
\label{Xmrelation2}
\end{equation}
for the contribution from two bosons cancel each other
in this case, which is due to
the relation $V_{11}=-V_{22}$.
Thus non-zero contribution are only from 
$\Gamma((\hX_m)^{2k}(\hX_{a})^{N_a}, D\!=\!2d)$ 
where $k=2,3,\ldots$ and $d=0,1,\ldots$.
The result is summarized as 
\begin{eqnarray}
\Gamma((\hX_m)^{N_m\ne 0}(\hX_a)^{N_a} , D=0)&=&
\Gamma((\hX_m)^{N_m\ne 0}(\hX_a)^{N_a} , D=0)_{B} 
\\
&&+
\Gamma((\hX_m)^{N_m\ne 0}(\hX_a)^{N_a} , D=0)_{C}
\; +O((\hX_i)^{7}).
\end{eqnarray}
where the first part contains no epsilon tensor
\begin{eqnarray}
\hspace{-1.3cm}\Gamma((\hX_m)^{N_m\ne 0}(\hX)^{N_a} , D=0)_B 
&=&-\frac{N_4}{8\lambda}\frac{1}{ r^3}
\int d\tau \str [\hX_{m_1},\hX_{m_2}]^2
\label{potm0}
\\
&& + \frac{3N_4}{8\lambda }\frac{r_a}{r^5}
\int d\tau \str ( \hX_{a}\,[\hX_{m_1},\hX_{m_2}]^2 )
\label{potm1}
\\
&& +\frac{3N_4}{16\lambda }\frac{1}{r^5}
\int d\tau \tr ( (\hX_{a})^2\,[\hX_{m_1},\hX_{m_2}]^2 )
\label{potm2A}
\\
&&-
\frac{15N_4}{16\lambda }\frac{r_{a_1} r_{a_2}}{r^7}
\int d\tau \str ( \hX_{a_1}\hX_{a_2}[\hX_{m_1},\hX_{m_2}]^2 )
\label{potm2B}
\\
&&
- \frac{N_4}{8 \lambda}\frac{1}{ r^5}
\int d\tau \tr ( [\hX_{m_1},\hX_{m_2}][\hX_{m_2},\hX_{m_3}] 
[\hX_{m_3},\hX_{m_1}])
\label{potmc}
\end{eqnarray}
and the second part has indices contracted using
epsilon tensor.
\begin{eqnarray}
&&\hspace*{-15mm}\Gamma((\hX_m)^{N_m\ne 0}(\hX_a)^{N_a} , D=0)_{C}
\nonumber\\
&=& +\frac{3N_4}{16 \lambda }\frac{r_a}{r^5}
\int d\tau \str ( \hX_{a}[\hX_{m_1},\hX_{m_2}][\hX_{m_3},\hX_{m_4}])
\epsilon_{ m_1m_2m_3m_4} 
\label{C61}
\\
&& +\frac{3N_4}{32 \lambda }\frac{1}{r^5}
\int d\tau \tr ( (\hX_{a})^2 [\hX_{m_1},\hX_{m_2}][\hX_{m_3},\hX_{m_4}])
\epsilon_{ m_1m_2m_3m_4} 
\label{C62A}
\\
&&
-\frac{15N_4}{32 \lambda }\frac{ r_{a_1} r_{a_2}}{r^7}
\int d\tau \str ( \hX_{a_1} \hX_{a_2} [\hX_{m_1},\hX_{m_2}][\hX_{m_3},\hX_{m_4}])
\epsilon_{ m_1m_2m_3m_4} 
\label{C62B}
\\
&& - \frac{N_4}{8 \lambda}\frac{1}{r^5}
\int d\tau \tr ( [\hX_{m},\hX_{m_1}][\hX_{m},\hX_{m_2}] [\hX_{m},\hX_{m_3}])
\epsilon_{mm_1m_2m_3}
\label{C62c}
\end{eqnarray}
where $\epsilon_{1234}=1$.

\subsection{Kinetic terms}
We summarize the results of terms with two derivatives.
For the terms with only $\hX_a$'s, 
\begin{eqnarray}
\Gamma((\hX_a)^{N_a},D=2 ) &=& -\frac{N_4\lambda}{4}\frac{1}{r^3}
\int d\tau \str (\dot{\hX}_{a}\dot{\hX}_{a})
\label{ki0}
\\
&&+\frac{3N_4\lambda}{4}\frac{r_a}{r^5}
\int d\tau \str (\hX_a\,\dot{\hX}_{b}\dot{\hX}_{b})
\label{ki1}
\\
& &+\frac{3 N_4\lambda}{8}\frac{1}{r^5}
\int d\tau \tr (\dot{\hX}_{a}\dot{\hX}_{a} \hX_{b}\hX_{b}) 
\label{ki2A}
\\ && -
\frac{15 N_4\lambda}{8}\frac{r_{a_1}r_{a_2}}{r^7}
\int d\tau \str ( \dot{\hX}_{a}\dot{\hX}_{a}\hX_{a_1}\hX_{a_2})
\label{ki2B}
\\
& &+
\frac{3 N_4\lambda}{8 }\frac{1}{r^5}
\int d\tau \tr ( \dot{\hX}_{a}\dot{\hX}_{b}[\hX_{b},\hX_{a}]) 
\label{kic1}
\\&&-
\frac{ N_4\lambda}{32 }\frac{1}{r^5}
\int d\tau \tr ( \partial_\tau[\hX_{a},\hX_{b}])^2
\label{kic2}
\\&&
\quad +O(\partial_{\tau}^2(\hX_a)^5 )
\nonumber
\end{eqnarray}
where the `dot' denotes derivative with $\tau$.
For the terms including  $\hX_m$, we have
\begin{eqnarray}
\hspace{-1cm}\Gamma((\hX_m)^{N_m\ne 0}(\hX_a)^{N_a} ,D=2 )
&=& 
\frac{N_4\lambda}{32 }\frac{1}{r^5}
\int d\tau \tr ( \partial_\tau[\hX_{m},\hX_{n}])^2
\label{kim1}\\
&&\hspace*{-1cm}+\frac{N_4\lambda}{64}\frac{1}{r^5}
\int d\tau \tr ( \partial_\tau[\hX_{m_1},\hX_{m_2}]
\partial_{\tau}[\hX_{m_3},\hX_{m_4}]) 
\epsilon_{m_1m_2m_3m_4}
\\&&
+O(\partial_{\tau}^2(\hX_i)^5 )\nonumber.
\end{eqnarray}
\subsection{Terms with one derivative}
Now we deal with terms with one derivative.
They only come from fermionic loops
$\Gamma^{\mbox{\scriptsize fermi}}$, and as a result, 
there are no contribution from $\hX_m$ for they do not
appear in the vertices for fermions.
The result for each number $N_a$ of $\hX_a$'s
is as follows.

The terms with $N_a\le 3$ vanish. 
For $N_a\ge 4$, we have
\begin{eqnarray}
\Gamma((\hX_a)^{4} ,D=1 )
&=&
\frac{3N_4}{8}\frac{r_a}{r^5}
\int d\tau \tr(\hX_{a_1}\hX_{a_2}\hX_{a_3}\dot{\hX}_{a_4})
\epsilon_{aa_1a_2a_3a_4}
\label{C41}
\\
\Gamma((\hX_a)^{5} ,D=1 )
&=&
\frac{N_4}{4}\frac{r_a}{r^7}
\int d\tau \tr(\hX_{a_1}\hX_{a_2}\hX_{a_3}\hX_{a_4}\dot{\hX}_{a_5})
\nonumber\\
&&\hspace*{-1.5cm}\times
\left(-2r_{a_4}\epsilon_{a_1a_2a_3a_5a}-2r_{a_1}\epsilon_{a_2a_3a_4a_5a}
+r_{a_2}\epsilon_{a_3a_4a_5a_1a}+r_{a_3}\epsilon_{a_2a_4a_5a_1a}
\right)
\label{C42}
\end{eqnarray}
\begin{eqnarray}
\Gamma((\hX_a)^{6} ,D=1 )
&=&
N_4\int d\tau \tr(\hX_{a_1}\hX_{a_2}\hX_{a_3}\hX_{a_4}\hX_{a_5}
\dot{\hX}_{a_6})
\nonumber\\
&&\quad\;\times
\Biggl\{
\frac{35}{48}\frac{r_a}{r^9} (
r_{a_1}r_{a_2}\epsilon_{a_3a_4a_5a_6a}+ r_{a_3}r_{a_4}\epsilon_{a_1a_2a_5a_6a}
+ r_{a_4}r_{a_5}\epsilon_{a_1a_2a_3a_6a}
\nonumber\\
&&\quad\quad + r_{a_1}r_{a_5}\epsilon_{a_2a_3a_4a_6a}
+ r_{a_2}r_{a_5}\epsilon_{a_1a_3a_4a_6a}+ r_{a_1}r_{a_3}\epsilon_{a_2a_4a_5a_6a}
)
\nonumber\\ 
&& \quad\quad -\frac{5}{16}\frac{r_{a_3}}{r^7}
\epsilon_{a_1a_2a_4a_5a_6}
\nonumber\\
&&\quad\quad
-\frac{5}{48}\frac{r_a}{r^7}
(
\delta^{a_1a_2}\epsilon_{a_3a_4a_5a_6a}+ \delta^{a_3a_4}\epsilon_{a_1a_2a_5a_6a}+
 \delta^{a_4a_5}\epsilon_{a_1a_2a_3a_6a}
\nonumber\\ 
&&\quad\quad+
 \delta^{a_1a_5}\epsilon_{a_2a_3a_4a_6a}+
\delta^{a_2a_5}\epsilon_{a_1a_3a_4a_6a}+ 
\delta^{a_1a_3}\epsilon_{a_2a_4a_5a_6a}
)
\biggl\}
\label{C43}
\\
&&-\frac{5N_4}{16}\frac{r_a}{r^7}
\int d\tau \tr(\hX_{a_1}\hX_{a_2}\hX_{a_3}\hX_{a_4}\hX_{a_5}
\dot{\hX}_{a_6})
\nonumber\\
&&\quad \times
(
\delta^{a_2a_3}\epsilon_{a_1a_4a_5a_6a}+  \delta^{a_4a_5}\epsilon_{a_1a_2a_3a_6a}-
 \delta^{a_1a_6}\epsilon_{a_2a_3a_4a_5a} 
\nonumber\\
&&\quad\quad-
\delta^{a_3a_5}\epsilon_{a_1a_2a_4a_6a}- 
\delta^{a_1a_3}\epsilon_{a_2a_4a_5a_6a}
).
\label{C4c}
\end{eqnarray}
Note that the last two terms of $\Gamma((\hX_a)^{6} ,D=1 )$
have the same index structures.
We have divided them for future purpose.



\section{Consistency with the Taylor and Van Raamsdonk's couplings}
The effective action of BD Matrix theory which was 
obtained in the previous section gives the 
D0-brane (Matrix theory) action in the background
of longitudinal 5-branes.  
We shall compare it with the couplings
to general weak background fields which was proposed by 
Taylor and Van Raamsdonk \cite{TR1}
from the analysis of BFSS Matrix theory. 
In the first subsection, we explain the proposal of 
ref.\cite{TR1} and show that BD Matrix theory
effective action is consistent with it, at the
leading order in the derivatives of backgrounds.
In the second subsection, subleading (higher-moment)
couplings are analyzed in detail. In the last subsection,
we discuss the implication of our result for the 
consistency between BD and BFSS matrix models.

\subsection{Agreement at the leading order}
We have obtained one-loop effective action for
the background matrices $X_i$ of the form
$(X_m, X_a) = (\hX_m ,r_a+\hX_a)$,
as an expansion in the time-derivatives and in $\hX_i$. 
As a result of the decomposition of $X_i$,
the background fields produced by the 5-branes
should appear as an expansion around a 
transverse position $r_a$.
We regard the following part of the effective action 
(rotated back to Minkowski signature) as the leading 
terms of the expansion.
\begin{eqnarray}
S&=& {1\over g_s\ell_s}\int dt \Tr \Big\{ 
{1\over 2}\dot{\hX}{}_m\dot{\hX}{}_m 
+{1\over 2}(1+{k\over r^3}) \dot{\hX}{}_a\dot{\hX}{}_a
+{1\over 4\lambda^2}(1-{k\over r^3})[\hX_m,\hX_n][\hX_m,\hX_n]\nonumber\\
&&
+{1\over 2\lambda^2}[\hX_a,\hX_m][\hX_a,\hX_m]
+{1\over 4\lambda^2}(1+{k\over r^3})[\hX_a,\hX_b][\hX_a,\hX_b]\Big\}
\nonumber\\
&&-i{3\over 4}{1\over g_s\ell_s}{kr_a\over \lambda r^5}
\epsilon_{a a_1 \ldots a_4}\int dt \Tr \{\hX_{a_1}\hX_{a_2}\hX_{a_3}
\dot{\hX}{}_{a_4}\}\nonumber\\
&&+{3\over 2}{1\over g_s\ell_s}{kr_a\over \lambda^2 r^5}
\epsilon_{m_1 \ldots m_4}\int dt \Tr \{\hX_{a}\hX_{m_1}
\hX_{m_2}\hX_{m_3} \hX_{m_4}\}
\label{matrixonL5}
\end{eqnarray}
where $k=\pi N_4 g_s \ell_s^3$. 

To discuss consistency with the general form of
couplings given in ref.\cite{TR1},
we first recall that the longitudinal 
5-brane solution is given as 
\begin{eqnarray}
ds^2&=& H^{-1/3}(-dx^0dx^0+dx^{10}dx^{10}+\eta_{{m}{n}}
dx^{{m}}dx^{{n}})
+ H^{2/3}\delta_{ab} dx^a dx^b \nonumber\\
F^{(4)}_{a_1\ldots a_4}&=&-\epsilon_{a_1\ldots a_4 a}
\partial_a H
\label{fivebr}
\end{eqnarray}
where $H$ is a harmonic function defined by
\begin{equation}
H=1+{k\over r^3}
\label{harmfn}
\end{equation}
with $r^2=(x^a)^2$. Field strength
$F^{(4)}$ is equivalently expressed by its dual
\[
F^{(7)}_{0m_1\ldots m_4 a 10}=-\epsilon_{m_1\ldots m_4}
\partial_a H^{-1}.
\]
The 5-brane is an electric source for the 
6-form potential.

Taylor and Van Raamsdonk's proposal for the Matrix theory
action in a weakly curved background is given as
follows \cite{TR1}.
\begin{eqnarray}
S&=&S_0+ \int dt \Big\{ 
{1\over 2}h_{MN}T^{MN}
+C^{(3)}_{MNP}J^{MNP} +C^{(6)}_{MNPQRS}M^{MNPQRS} \Big\}\nonumber\\
&&+\sum_{n=1}^{\infty}{1\over n!} \int dt 
\Big\{ {1\over 2}
\partial_{i_1}\ldots\partial_{i_n}h_{MN}T^{MN;i_1\ldots i_n}
+\partial_{i_1}\ldots\partial_{i_n}C^{(3)}_{MNP}
J^{MNP;i_1\ldots i_n}
\nonumber\\
&&+\partial_{i_1}\ldots\partial_{i_n}C^{(6)}_{MNPQRS}
M^{MNPQRS;i_1\ldots i_n} \Big\}
\label{linearized}
\end{eqnarray}
where $M,N=0,1,\dots,10$ are the 11D indices. 
Here, $T^{MN}$, $J^{MNP}$ and 
$M^{MNPQRS}$ are the energy-momentum 
tensor, `membrane current' and `5-brane current' of Matrix theory,
respectively. They were identified by interpreting the one-loop effective
potential between two diagonal blocks in BFSS Matrix theory as
the result of 
tree-level interaction of DLCQ supergravity \cite{KT,TR1}.
Explicit forms of the bosonic part 
for the components which are needed for our discussion
are
\begin{eqnarray}
T^{+-}&=&{1\over g_s\ell_s}\str \left({1\over 2}\dot{\hX}_i\dot{\hX}_i
-{1\over 4\lambda^2}[\hX_i,\hX_j][\hX_i,\hX_j]\right)\nonumber\\
T^{ij}&=&{1\over g_s\ell_s}\str \left(\dot{\hX}_i\dot{\hX}_j
-{1\over \lambda^2}[\hX_i,\hX_k][\hX_k,\hX_j]\right)\nonumber\\
J^{a_1 a_2 a_3}&=&{-i\over 6g_s\ell_s\lambda}\str 
\left( \dot{\hX}_{a_1} [\hX_{a_2},\hX_{a_3}] +\dot{\hX}_{a_2} [\hX_{a_3},\hX_{a_1}]
+\dot{\hX}_{a_3} [\hX_{a_1},\hX_{a_2}]\right)\nonumber\\
M^{+- m_1 \ldots m_4}&=&{-1\over 720 g_s\ell_s\lambda^2}\str 
\left([\hX_{m_1},\hX_{m_2}][\hX_{m_3},\hX_{m_4}] 
+ [\hX_{m_1},\hX_{m_3}][\hX_{m_4},\hX_{m_2}] \right. \nonumber\\
&&\left.  +[\hX_{m_1},\hX_{m_4}][\hX_{m_2},\hX_{m_3}]\right) 
\label{currents}
\end{eqnarray}
where we have changed the 
sign of $J^{a_1 a_2 a_3}$ and 
the coefficient of $M^{+- m_1 \ldots m_4}$ 
from the ones of refs.\cite{KT, TR1} 
to adjust to normalization of the antisymmetric tensor fields.

The last two lines of eq.(\ref{linearized}) state that
$n$-th derivative of background fields should couple to  
the matrix version of $n$-th moment of the currents.
$n$-th moment is given from the above operators
as follows.
\EQA
T^{MN;i_1\ldots i_n}&=&{\rm Sym}(T^{MN};\hX_{i_1},\ldots,\hX_{i_n})
\nonumber\\
J^{MNP;i_1\ldots i_n}&=&{\rm Sym}(J^{MNP};\hX_{i_1},\ldots,\hX_{i_n})
\nonumber\\
M^{MNPQRS;i_1\ldots i_n}&=&{\rm Sym}
(M^{MNPQRS};\hX_{i_1},\ldots,\hX_{i_n})
\EQN
where the RHS means the symmetrized trace is taken 
after inserting $\hX_i$ $n$ times into 
the expressions inside the trace.
(When symmetrizing the ordering, 
 $[\hX_i,\hX_j]$ are treated as a single unit, as in the
previous section.)
We can also say that background fields enter the action as 
`non-abelian Taylor expansion' around some point $x_i=r_i$,
where the coordinates in the series are replaced by matrices:
\begin{equation}
\varphi = \sum_{n=0}^{\infty}\frac{1}{n!}
\hX_{i_1}\cdots \hX_{i_n} 
(\partial_{i_1}\cdots\partial_{i_n})
\varphi\bigg|_{x_i=r_i}.
\label{NAT}
\end{equation}

The proposed couplings to $C^{(3)}_{a_1a_2a_3}$
and $C^{(6)}_{+-i_1i_2i_3i_4}$ at the first few orders
can be rewritten as 
\begin{eqnarray}
&&\sum_{n=0}^3{1\over n!}\partial_{i_1}\ldots\partial_{i_n}
C^{(3)}_{a_1a_2a_3}J^{a_1a_2a_3;i_1\ldots i_n}=
- \frac{i}{4 g_s\ell_s\lambda}{\rm Tr}
(\hX_{a_1} \hX_{a_2} \hX_{a_3} \dot{\hX}_{a_4}) 
F^{(4)}_{a_1a_2a_3a_4}
\nonumber\\
&&\hspace*{2.5cm}- \frac{i}{30 g_s\ell_s\lambda}{\rm Tr}
(\hX_{a_1} \hX_{a_2} \hX_{a_3} \hX_{a_4} \dot{\hX}_{a_5})
\left[4 \partial_{(a_1}F^{(4)}_{a_4)a_2a_3a_5} -
2 \partial_{(a_2}F^{(4)}_{a_3)a_4a_5a_1}
\right]
\nonumber\\
&&\hspace*{2.5cm} -\frac{i}{36g_s\ell_s\lambda}{\rm Tr}
(\hX_{a_1} \hX_{a_2} \hX_{a_3} \hX_{a_4} \hX_{a_5}\dot{\hX}_{a_6})
\nonumber\\
&&\hspace*{3cm}\quad\times
\left[\partial_{a_1}\partial_{(a_2}F^{(4)}_{a_3)a_4a_5a_6} 
+\partial_{a_4} \partial_{(a_3}F^{(4)}_{a_5)a_1a_2a_6} +
\partial_{a_5} \partial_{(a_1}F^{(4)}_{a_2)a_3a_4a_6}
\right]  \, ,\label{CS3}\\
%
&&\sum_{n=0}^2{1\over n!}\partial_{i_1}\ldots\partial_{i_n}
C^{(6)}_{+-i_1i_2i_3i_4}M^{+-i_1i_2i_3i_4;i_1\ldots i_n} =
\frac{-1}{10 g_s\ell_s\lambda^2}{\rm Tr}
(\hX_{i_1} \hX_{i_2} \hX_{i_3} \hX_{i_4} \hX_{i_5})
F^{(7)}_{+-i_1i_2i_3i_4i_5}
\nonumber\\
&&\hspace{2.5cm}-\frac{1}{12 g_s\ell_s\lambda^2}{\rm Tr}
(\hX_{i_1} \hX_{i_2} \hX_{i_3} \hX_{i_4} \hX_{i_5}\hX_{i_6})
\partial_{i_6} F^{(7)}_{+- i_1i_2i_3i_4i_5} 
\quad 
\label{CS5}
\end{eqnarray}
where we have used the partial integration and the cyclic
symmetry of the trace. 
Note that the zeroth moment
$J^{a_1 a_2 a_3}$ is a total 
derivative and  
$M^{+- m_1 \ldots m_4}$ vanishes for the cyclic symmetry of the trace,
thus, the leading contributions are from $J^{a_1 a_2 a_3; a_4}$ and
$M^{+- m_1 \ldots m_4;a}$. 
Also note that the terms involving the derivatives of field strengths,
different expressions are also possible.


We can see that the part (\ref{matrixonL5}) of the BD Matrix 
theory effective action
precisely agree with the lowest-moment contribution of 
eq.(\ref{linearized})
\[
S=S_0+ \int dt \Big\{ {1\over 2}h_{MN}T^{MN}
+J^{a_1 a_2 a_3; a_4}\partial_{a_4}
C^{(3)}_{a_1 a_2 a_3} 
+30M^{+- m_1 \ldots m_4;a}\partial_a
C^{(6)}_{+- m_1 \ldots m_4} \Big\}
\]
upon substitution of the longitudinal 5-brane background
at the linear order
\begin{eqnarray*}
&&h_{+-}= {1\over 3}{k\over r^3},\; 
h_{mn}= -{1\over 3}{k\over r^3}\delta_{mn},\;
h_{ab}= {2\over 3}{k\over r^3}\delta_{ab}\\
&&
F^{(4)}_{a_1 a_2 a_3 a_4}= 3{kr_a\over r^5}\epsilon_{a_1 a_2 a_3 a_4 a},\;
F^{(7)}_{+- m_1 m_2 m_3 m_4 a}=-3{kr_a \over r^5}\epsilon_{m_1 m_2 m_3 m_4}.
\end{eqnarray*}


\subsection{Subtleties for the higher-moment couplings}
Now, we shall examine the terms which are of higher orders in $\hX_i$.
First, let us consider the coupling with the 11D metric.
According to the Taylor and Van Raamsdonk's proposal, 
the coupling is given from the first two lines of (\ref{matrixonL5})
by replacing $k/r^3$ with the non-abelian Taylor expansion
\[
{k\over r^3}\rightarrow \frac{k}{r^3}-\frac{3k}{r^5}r_a\hX_a
+\frac{k}{2}
\left(
-\frac{3}{r^5}\hX_a\hX_a +15\frac{r_a r_b}{r^7}\hX_a\hX_b
\right)+\cdots \,.
\label{NATaylorr3}
\]
This part  
is to be compared with the part of the BD effective action 
$\Gamma((\hX_a)^{N_a},D=0)$, 
$\Gamma((\hX_m)^{N_m} (\hX_a)^{N_a},D=0)_B$ 
and $\Gamma((\hX_i)^{N_i},D=2)$ which we have obtained in 
sections 3.2 and 3.3.
Exact agreement goes through to the subleading order for this case.
The terms (\ref{pota1}), (\ref{potm1}) and (\ref{ki1})
in our result agree with the proposed first moment-couplings 
(eq.(\ref{matrixonL5}) evaluated at the 
first order of non-abelian Taylor expansions).
However, we find discrepancies at the next order.
The terms 
 (\ref{pota2A}), (\ref{pota2B}), (\ref{potm2A}), (\ref{potm2B}),
(\ref{ki2A}) and (\ref{ki2B}),
in BD effective action have the coefficients expected
from the proposed second-moment couplings, however,
there are subtleties in the ordering of matrices.
Some of the terms  
((\ref{pota2B}), (\ref{potm2B}) and (\ref{ki2B}))
satisfy the symmetrized trace prescription,
but others ((\ref{pota2A}), (\ref{potm2A}) and (\ref{ki2A})) do not.
Moreover, our result has corrections involving extra commutator of
matrices (\ref{potac}), (\ref{potmc}), (\ref{kic1}) and (\ref{kic2}), 
which are not present in the proposed action.

Next, we analyze the couplings to the 3-form potential.
The $\Gamma((\hX_a)^{N_a},D=1)$ part of the BD effective 
action obtained in section 3.4 represent these couplings.
Exact agreement holds at the subleading order, also for
this case.
The term (\ref{C42}) agree with the
second term of eq.(\ref{CS3}).
At the next order (third-moment coupling), the proposed 
coupling (\ref{CS3})
agrees with a part (\ref{C43}) of the BD effective action,
however, the latter has  extra contribution (\ref{C4c}).

The couplings to the 6-form potential are given by the
$\Gamma((\hX_m)^{N_m}(\hX_a)^{N_a},D=0)_C$ part
in the BD Matrix theory effective action.
For these couplings, we find
disagreement already at the subleading order (second-moment
coupling). Substituting the background into 
the second term of eq.(\ref{CS5}), proposed coupling reads 
\[
\frac{3N_4}{32\lambda}
\left(\delta_{ab}\frac{1}{r^5} -5\frac{r_ar_b}{r^7}
\right)
\str ([\hX_{m_1}, \hX_{m_2}][\hX_{m_3},\hX_{m_4}]\hX_a\hX_b)
\epsilon_{m_1m_2m_3m_4}.
\label{CS5l}
\]
The terms (\ref{C62A}) and (\ref{C62B}) in the BD effective
action have the same coefficients as the first and the second
terms of the above expression, respectively. However, there 
is a difference
in the ordering for (\ref{C62A}) ($\Tr$, not $\str$).
Our result also have 
a correction (\ref{C62c}) containing extra $\hX_m$'s in the
form of commutators.  

Finally, BD Matrix theory effective action
have corrections to the kinetic terms of $\hX_m$ 
such as (\ref{kim1}), which
is not expected from the proposed action, 
also in the form of commutators.

\subsection{On the consistency between BFSS 
and BD matrix models}
As described in the previous subsections,
we have confirmed that the couplings
which are expected from the proposal of Taylor and
Van Raamsdonk indeed exists in the effective action
of BD Matrix theory. We also found subtleties in the ordering
of matrices, that is, there are corrections to the
above proposal involving extra commutators.

To discuss consistency between BFSS and BD matrix models,
we first mention that the ordering problems similar
to the ones which we have found is also present
in the effective action of BFSS Matrix theory. 
In ref.\cite{KT}, effective action was obtained
using the `quasi-static approximation' including the 
contributions of the higher moments of arbitrary orders.
(In ref.\cite{TR1}, which uses the similar method of approximation as 
ours, only the zeroth and first moments were calculated explicitly.)
It was noted in ref.\cite{KT}, that
the effective action is not given by
the symmetrized trace in the usual sense, 
when there is a contraction of indices between two of the matrices 
which had been inserted to construct the moments. 
This is precisely the same situation as the one for 
the terms such as (\ref{pota2A}), (\ref{potm2A}) and (\ref{ki2A}).
In the BFSS effective action, some of the terms will be
interpreted as interactions between two (block diagonal) 
symmetrized trace operators but others will not.

The agreement of the symmetrized trace part
of the BD effective action with the Taylor and 
Van Raamsdonk's couplings 
 suggests the consistency between the two
matrix models.
To examine the consistency between BFSS and BD matrix models further,
it must be helpful to extend the analysis of ref.\cite{TR1} 
to the next order and study the contribution from the
second moment operators to the BFSS effective action.

\section{Interpretation from the 10D perspective}
Before going to the conclusion, we shall briefly discuss
our results from the 10D perspective. The action of
BD Matrix theory is the SYM for the D0-D4 system, which
consists of only the lowest modes of open strings.
Thus, the effective action resulting from the integration
of $v$ and $\chi$ (lowest modes of 0-4 strings) is 
guaranteed to be valid
when their mass are smaller than the mass of the higher
modes of open strings, that is, when the distance between
D0-branes and D4-branes is smaller than the string scale
($r\ll \ell_s$).
If we want to discuss the long distance interaction, 
the open-string cylinder amplitudes between 
D0-brane and D4-brane must be 
studied.\footnote{It is a well-known fact that the SYM result for
the $(\partial_t\hX_a)^2$-term in the abelian case is valid for
$r\gg\ell_s$ as well. Contributions from massive modes
of D0-D4 open strings cancel for this term \cite{DKPS}.} 

In this section we compare BD Matrix theory effective
action with a proposal for the D-brane action 
in curved space due to Myers \cite{My},
and point out some agreement. 
We must emphasize that the region of validity
of the two action is different, for 
we cannot expect that the Born-Infeld like action 
introduced in the following is valid at
short distance.

Myers proposed the following form of
the D-brane action in curved space, motivated by the consistency
with a single D9-brane action when the T-duality invariance
is assumed \cite{My}. For D0-branes, it reads
\[
S_{\mbox{\scriptsize Myers}}=
\int dt ({\cal L}_{{\rm BI}}+{\cal L}_{{\rm CS}}^{(3)} 
+{\cal L}_{{\rm CS}}^{(5)})
\]
where the Born-Infeld (BI) part is given as
\begin{equation}
{\cal L}_{{\rm BI}} =-{1\over g_s\ell_s} \str 
\left(e^{-\phi}
\sqrt{-(E_{00}+
(Q^{-1})^j{}_k D_t{X}^i D_t{X}^k E_{ij})}
\sqrt{
\det(Q^j{}_k)
}\right)
\label{LBI}
\end{equation}
where  
\[
E_{\mu\nu}=G_{\mu\nu}+B_{\mu\nu},\;\;
Q^{j}{}_{k}= \delta^j{}_k +i \frac{1}{\lambda} [X^j,X^i]E_{ik}.
\]
The Chern-Simons (CS) terms which are relevant to the
coupling with D4-branes are given as
\begin{eqnarray}
\hspace{-1cm}
{\cal L}^{(3)}_{{\rm CS}} &=& \frac{i}{2 g_s\ell_s\lambda}
{\rm STr}\left(C^{(3)}_{ijk}[X^k,X^j]\partial_tX^i
\right),
\label{l3}
\\
{\cal L}^{(5)}_{{\rm CS}} &=& -\frac{1}{8 g_s\ell_s\lambda^2}
{\rm STr}\left(
C^{(5)}_{0ijkl}[X^j,X^i][X^l,X^k]
\right)
.
\label{l5}
\end{eqnarray}
In the above action,  background fields are prescribed to be 
given as the non-abelian Taylor expansion (\ref{NAT}).
Also note that the static gauge $x^0=t$ is assumed.

We consider the part
of this action which is the leading terms 
in $\alpha'\rightarrow 0$ limit when $\hX_i/\alpha'$ and
the background (including the series of the
non-abelian Taylor expansion) are 
fixed.\footnote{This part is the one which allow an interpretation
in terms of DLCQ M-theory. Following the argument of Seiberg and Sen,
DLCQ M-theory is given from 10D type IIA string theory
by an infinite boost in the compactified
11-th direction with a rescaling of the length scale by 
an infinite factor. (See ref.\cite{TR2} for explicit
transformation rules.) This part of D0-brane action 
remain non-vanishing after the transformation.}
Substituting the D$4$-brane solution in the string frame 
metric 
\begin{eqnarray*}
&&ds^2 = H^{-1/2}
(-dx^0 dx^0+\delta_{mn}dx^{m} dx^{n})
+H^{1/2}\delta_{ab}dx^{a} dx^{b},\;\;
e^{\phi} = H^{-1/4}\nonumber\\
&&F^{(6)}_{0m_1m_2m_3m_4 a}= 
\epsilon_{m_1m_2m_3m_4} \partial_{a}(H^{-1}),\quad
F^{(4)}_{a_1a_2a_3a_4}=-\epsilon_{ a_1a_2a_3a_4a}\partial_{a}H 
\label{bgF6},\quad H=1+{k\over r^{3}}
\end{eqnarray*}
into the action, the $\alpha'$-leading part is given
as follows. The Born-Infeld part reduces to
\begin{eqnarray}
S_{{\rm BI}}&\rightarrow&T_0\int dt \str\left\{
\frac{1}{2}\dot{\hX}_m\dot{\hX}_m 
+\frac{1}{2}H\dot{\hX}_a\dot{\hX}_a
\right.
\nonumber\\
&& \left.
+\frac{1}{\lambda^2}
\left(\frac{1}{4}H^{-1}[\hX_m,\hX_n]^2
+\frac{1}{2}[\hX_m,\hX_a]^2
+\frac{1}{4}H[\hX_a,\hX_b]^2
\right)
\right\}
\label{BIMy}
\end{eqnarray}
where we have taken the $A_0=0$ gauge. 
The CS terms (\ref{l3}) and (\ref{l5}) are of the same order as
eq.(\ref{BIMy}).

At the linearized level of the background (up to
the part linear in $k\propto N_4$), the terms
(\ref{l3}), (\ref{l3}) and (\ref{BIMy}) agree
with the Taylor and Van Raamsdonk's action 
as we see from eqs.(\ref{linearized}), (\ref{currents}),
(\ref{CS3}) and (\ref{CS5}). 
Thus the consistency (up to subtleties in the ordering
of matrices) of our result with Myers action
is as explained in section 4.
There, we pointed out the existence of the terms
which do not satisfy symmetrized trace prescription
in the Matrix theory effective action. 
It will be appropriate to mention here a
possibility for a difference in the
long distance behavior depending on the 
ordering of matrices: It may be the case that only
terms with the symmetrized ordering are protected
by SUSY and are allowed to be interpreted
as  long distance interactions. Of course, to test
the above statement, we must develop non-renormalization
theorems in the gauge theory or study cylinder amplitudes 
between the multiple branes. These are important issues
for the future study.

In addition, Myers' action predicts that there are 
corrections for $[\hX_m,\hX_n]^2$ term and for the coupling
to 5-form (6-form in 11D) potential coming 
from the expansion of $H^{-1}$ in $k$.
These non-linear effects of the  background fields
are expected to be reproduced in BD Matrix theory by 
higher-loop effects.

\section{Discussions}
In this paper, we studied one-loop effective action of 
Berkooz-Douglas Matrix theory.
BD Matrix theory is a proposal
for the definition of M-theory in the presence
of longitudinal 5-branes, and it has extra degrees of
freedom (which are the lowest modes of D0-D4 string)
compared to the original BFSS Matrix theory.
The result of integrating out the extra fields gives 
the effective action for the D0-brane degrees of freedom
in the background field produced by
the 5-branes.

Since the 5-branes which we are dealing with have
no D0-brane charge and cannot be
realized in the BFSS Matrix theory, our analysis
provides a non-trivial check for the coupling of D0-branes
to general weak background fields  proposed by
Taylor and Van Raamsdonk \cite{TR1,TR2} from the
Matrix theory analysis.  
We have confirmed that the couplings given by
inserting the longitudinal 5-brane solution (at the linearized
level) in the
above proposal appear in the effective action of 
BD Matrix theory precisely as expected.
We also found that there are corrections
involving extra commutators, which violates
the proposed symmetrized trace prescription.
However, the similar subtleties for the ordering
of matrices are also present in the effective action of 
BFSS Matrix theory itself. 
The exact agreement between the symmetrized trace terms
in the BD effective action 
and the Taylor and Van Raamsdonk's proposal
can be regarded as a suggestion for the 
consistency between the BD and BFSS matrix models.
To discuss the consistency in more detail, it should be
helpful to extend the analysis of the BFSS effective action
performed in ref.\cite{TR1} to the higher order terms
(second moment couplings).

We think that the coupling of matrix fields
to the  background supergravity fields, and in particular,
the ordering problem for the product of matrices are issues
which need further analyses.
Most direct way to obtain the action of D-branes
should be the calculation of string scattering amplitudes.
(See refs.\cite{GM1,GM2,OO} among others.)
It may be necessary to perform thorough study of the 
scattering amplitudes for the operators of all possible orderings
at the order of interest.

Finally, we shall list other problems to be studied.

1) Direct extension of this work is the study of
the fermionic part of the one-loop effective action.
It allows the following interesting consistency
check of the action which was originally proposed
in ref.\cite{TR2}: By interpreting the effective action 
as the Matrix theory action on the longitudinal 
5-brane background, we may evaluate the 
quantum effective action for two diagonal blocks,
for example, starting from that action. 
It should correspond to the supergravity interaction 
between two objects evaluated on that background.

In addition, it is an important problem to find a 
gauge invariant ($\kappa$-symmetric) action of 
multiple D-branes which reduces to Matrix theory 
effective action after gauge fixing. 
In ref.\cite{B1}, ambitious attempt was made
to define a generalization of $\kappa$-symmetry
which have non-Abelian parameters. However, the
construction does not seem successful, for 
the action is not consistent with string 
amplitudes \cite{B2}. 
It may be more suitable to introduce 
$\kappa$-symmetry only for the U(1) (center of mass) 
part, as in ref.\cite{So}.

2) Studies toward establishing the validity of
BD Matrix theory as a fundamental theory are 
definitely important. Firstly, cylinder amplitudes
between D0-branes and D4-branes should be analyzed in
detail. We want to clarify in what cases open string 
massive modes cancel and the SYM is able to describe
long distance physics. Also, most interesting problem
is whether BD Matrix theory can reproduce the non-linear
supergravity fields of the 5-branes, as mentioned
in section 5. Those kinds of analyses for this version 
of Matrix theory
with half the maximal SUSY will give implications
on the connection between matrix models and gravity.

3) An interesting physical phenomenon which is expected
to occur in the D4-brane background is the so-called
Myers effect. As we have seen, multiple D0-branes
can couple to the 4-form field strength produced
by D4-brane. This coupling should give rise to a 
stable non-commutative configuration of D0-branes which has
the shape of 2-sphere, following the qualitative argument
first done by Myers \cite{My}. In ref.\cite{A}, one
of the present authors performed a detailed analysis of this
problem. It was shown that a certain spherical configuration 
is indeed a solution of the equation of motion of
the Myers' action for D0-branes in the D4-brane background,
by taking a special coordinate system where some of the coordinates  
 are assumed to be commutative. 
Furthermore, a spherical configuration which exhibits
exactly the same kinematical behavior as the point-like
D0-branes was found.
Similar result is likely to be reached
in the framework of BD Matrix theory.
However, the effective action which was obtained in this paper 
is not suitable for describing the configuration 
studied in ref.\cite{A}, for it will require all orders
of the expansion in the derivatives of background fields. 
We hope to study quantum corrections 
around the background matrices of the form of the
configuration of ref.\cite{A} directly and discuss its stability
in BD Matrix theory. 
(In a recent paper \cite{Ra}, non-commutative configurations
of D0-branes with open topology ending on D4-branes
were studied in BD Matrix theory coupled to external
supergravity fields. What we mean here is configurations with
closed topology in the theory without additional
external fields.)
The search for a finite-sized stable configuration 
of a collection of $N$ fundamental degrees of freedom
must be important in regards of its possible connection 
to the holographic principle.

\vspace{0.4cm}
\paragraph{Acknowledgements}

The authors would like to thank T. Yoneya for clarifying comments,
and T. Muramatsu for helpful discussions.
They also thank D. Sorokin for a comment on the
$\kappa$-symmetry of multiple D-branes.

\end{document}